\documentclass[nopubldata]{lpor2014}
\usepackage{blindtext}
\usepackage{hyperref}
\usepackage{amsmath}
\usepackage{graphicx}
\usepackage{color}
\usepackage{wrapfig}
\usepackage{epstopdf}
\hypersetup{%
 colorlinks,
 plainpages=false,
 pdfpagelabels,
 breaklinks,
 pdfview=Fit,
 bookmarksopen,
 bookmarksnumbered,
 linkcolor=black,
 anchorcolor=black,
 citecolor=black,
 filecolor=black,
 urlcolor=LPRred
 }%
\graphicspath{{./figs/}}
\setpages[]{}
\setvolume[0]{0}
\setyear{2011}%
\setdoi{201100000}%
\oddheadlogotext{}{Original Paper}
\begin{document}

\titlefigure{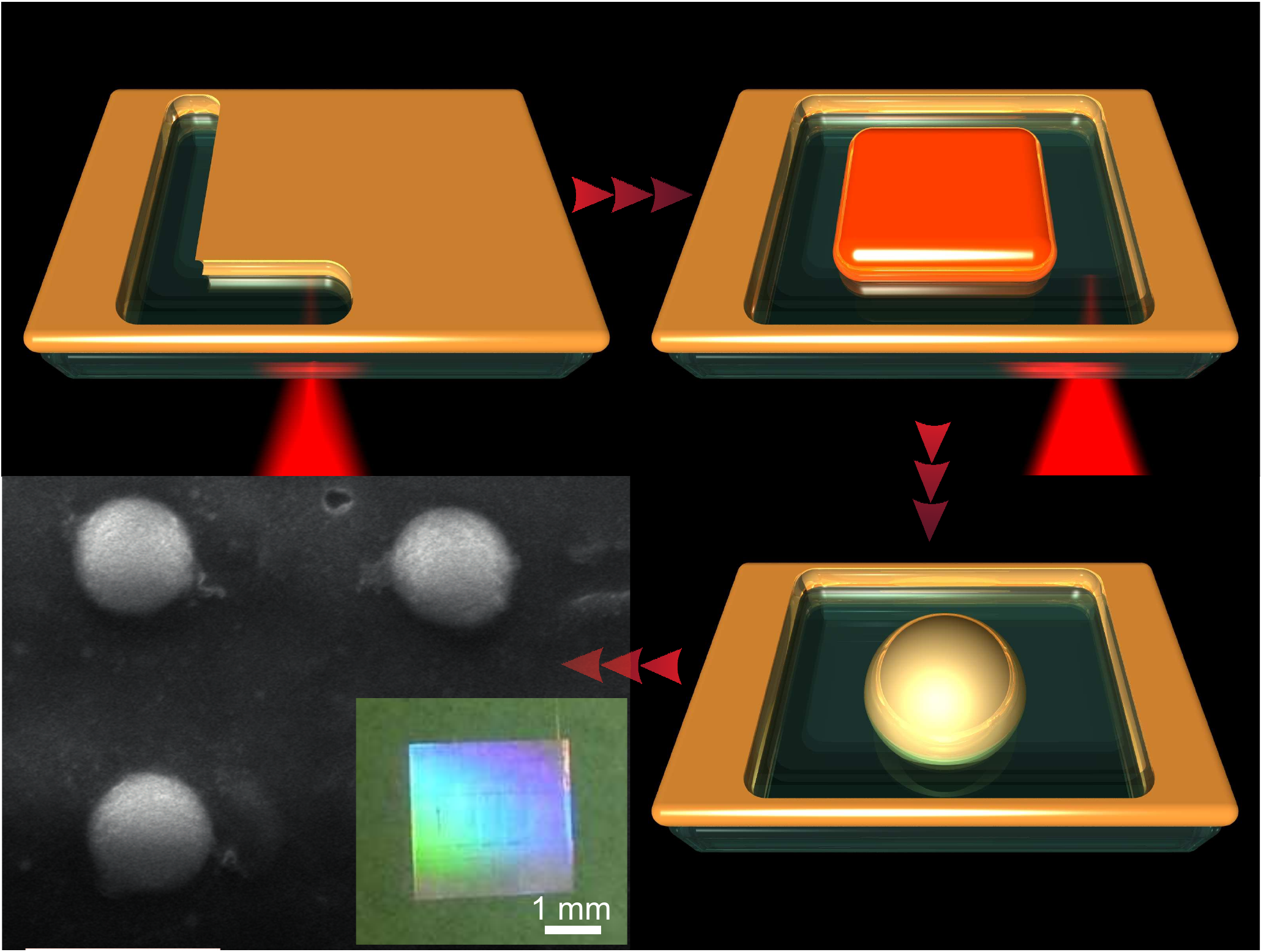}
\abstract{
Dewetting of thin metal films is one of the most widespread method for functional plasmonic nanostructures fabrication. However, simple thermal-induced dewetting does not allow to control degree of nanostructures order without additional lithographic process steps. Here we propose a novel method for lithography-free and large-scale fabrication of plasmonic nanostructures via controllable femtosecond laser-induced dewetting. The method is based on femtosecond laser surface pattering of a thin film followed by a nanoscale hydrodynamical instability, which is found to be very controllable under specific irradiation conditions. We achieve control over degree of nanostructures order by changing laser irradiation parametrs and film thickness. This allowed us to exploit the method for the broad range of applications: resonant light absorbtion and scattering, sensing, and potential improving of thin-film solar cells.}

\title{Controllable Femtosecond Laser-Induced Dewetting for Plasmonic Applications}

\author{Sergey~V.~Makarov\inst{1,*}, Valentin~A.~Milichko\inst{1}, Ivan~S.~Mukhin\inst{1,2}, Ivan~I.~Shishkin\inst{1}, Dmitriy~A.~Zuev\inst{1}, Alexey~M.~Mozharov\inst{1,2}, Alexander~E.~Krasnok\inst{1}, and Pavel~A.~Belov\inst{1}}%
\authorrunning{Sergey V. Makarov et al.}
\mail{\email{s.makarov@metalab.ifmo.ru}}

\institute{
Departament of Nanophotonics and Metamaterials, ITMO University, St.~Petersburg
\and
Laboratory of Renewable Energy Sources, St. Petersburg Academic University, St.~Petersburg}


\maketitle

\section{Introduction}

\begin{figure*}
\begin{center}
\includegraphics[width=0.95\textwidth]{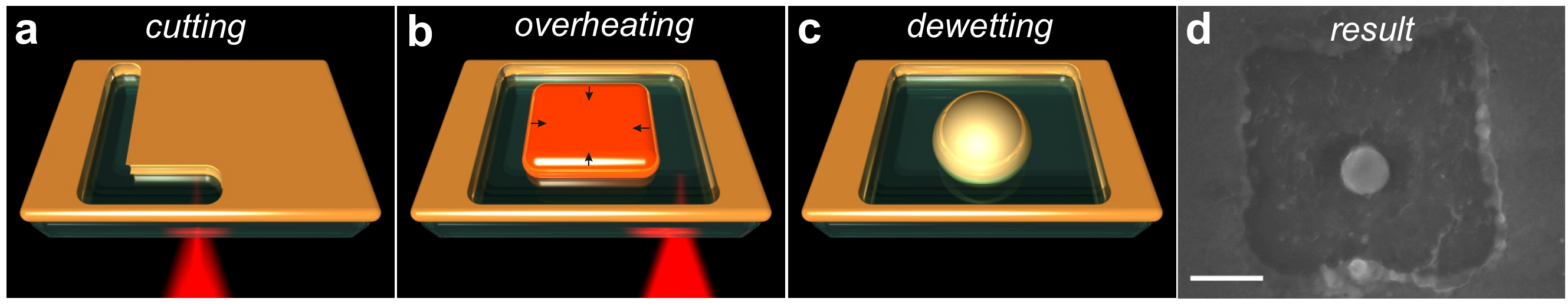}
\end{center}
\caption{Schematic illustration of single nanoparticle formation under femtosecond laser cutting of patch from a Au film on a dielectric substrate (a-c). (d) SEM image of a single Au nanoparticle on a SiO$_{2}$ substrate fabricated by fs-laser irradiation of a 30-nm Au film at the fluence of 40~mJ/cm$^{2}$. Scale bar is 500~nm.}\label{scheme}
\end{figure*}

The resonant metallic nanoparticles are proven to be efficient systems for the electromagnetic field control at nanoscale, owing to the ability to localize and enhance optical field via excitation of strong plasmon resonances~\cite{maier2007plasmonics, luk2013fano}. Plasmonic devices requiring ordered nanoparticles are metasurfaces~\cite{meta2014capasso, alu2014giant, meta2015kivshar}, waveguides~\cite{maier2003local}, and plasmonic filters~\cite{grigorenko08extremely, boltasseva2008plasmonic}; whereas disordered nanoparticles are used mostly in broadband plasmonic devises for biosensing~\cite{moskovits1985surface, Halas2007nano, anker2008biosensing} and photovoltaics~\cite{atwater2010plasmonics, cui2014plasmonic}.

To date, the most popular and controllable approaches of plasmonic nanostructures fabrication are based on direct ion-beam milling or multistage e-beam and nanoimprint lithographes. However, lithography-free and single-step methods are more desirable for both ordered and disordered large-scale nanostructuring.~Among single-step techniques, chemical syntheses of monodisperse nanoparticles colloid is a promising method for high-throughput fabrication of disordered structures and requires additional technological steps to order the nanoparticles into functional nanostructure~\cite{Junno95,print2007nanoparticle,Shi2013NatCom, patra2014plasmofluidic}. Nanoparticles fabrication via laser-induced transfer from thin films~\cite{kuznetsov2009, chichkov2014NatCom} and lithographically prepared structures~\cite{kuznetsov2011} placed on a certain distance from a receiving substrate is another way for controllable fabrication of plasmonic nanostructures.

Despite many different approaches have been proposed dewetting of thin metal films remains one of the most widespread method for functional plasmonic nanostructures fabrication. Indeed, a self-assembly process via dewetting of heated thin film is an cost-effective and environment-friendly approach for almost disordered nanostructures fabrication. Laser radiation~\cite{Bischof96,Trice07,fowlkes2011self,Wu11}, ion-beam~\cite{Lian06} or heating on a hotplate~\cite{dewetting2000anisotropic,Kim09APL,Bobod2014wafer} have been utilized as heat sources for launching the spontaneous dewetting process. In turn, controllable fabrication of ordered metal nanparticles arrays has been achieved via expensive and time-consuming lithography-assisted dewetting~\cite{Wu11, Lian06, Kim09APL,fowlkes2011self,wu2014directed}. Therefore, development of a method based on lithography-free and single-step dewetting, making possible fabrication of both ordered and disordered nanostructures, would be very prospective for a broad range of plasmonic applications.

In this work we present single-step, lithography-free, and cost-efficient method for large-scale fabrication of ordered and disordered plasmonic nanostructures. This method unifies the simplicity of thin film dewetting process with precision and high-productivity of the laser technology, enabling extreme simplification of the nanoparticles fabrication process, while the quality of fabricated structure is well-maintained. In other words, we achieve single-step controllable dewetting by strongly focused femtosecond laser pulses, where the laser irradiation is used to pattern and heat-up the residual film to the dewetting temperatures.~We implement this simple method for writing of 0D, 1D, 2D plasmonic (several types) structures, and show their applicability for resonant absorbtion/scattering of light, sensing, and photovoltaics.

\section{Principles of fabrication and optimization}

In order to make the method of Au nanoparticles array fabrication as simple as possible, we replace all lithographical stages of film patterning~\cite{Kim09APL, fowlkes2011self, ye2011templated, fowlkes2014hierarchical} by their direct fs-laser cutting with simultaneous heating of the remain patches up to the dewetting temperatures (Fig.~\ref{scheme}, for details of fabrication, see \textit{Methods} and Fig.S1 in
\textit{Supplementary Materials}). Although the proposed method is single-step, it is divided into several physical stages.

\begin{figure*}[!t]
\begin{center}
\includegraphics[width=0.85\textwidth]{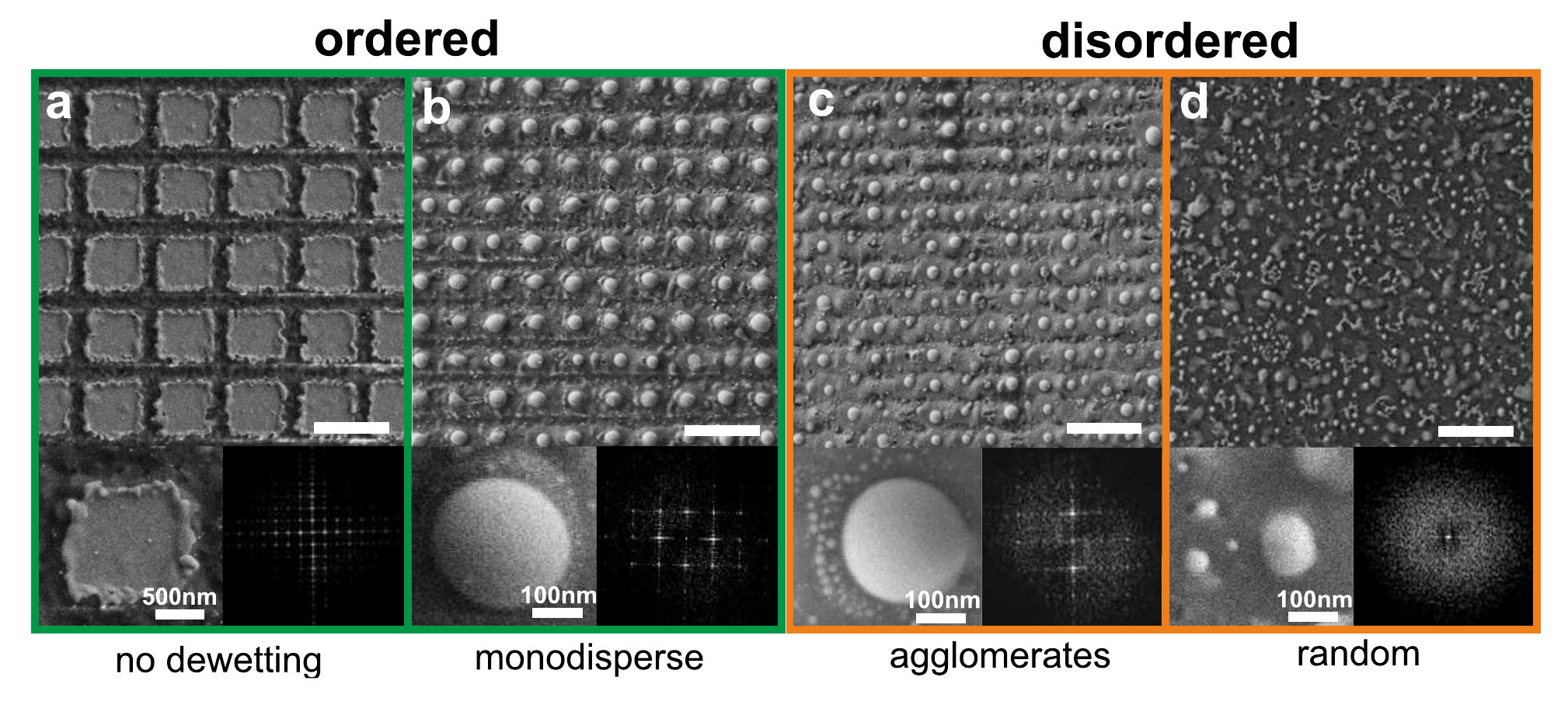}
\end{center}
\caption{SEM images of Au nanoparticles arrays fabricated from a 30-nm film on a SiO$_{2}$ substrate at the fluence of 40~mJ/cm$^{2}$ and different periods of laser scan: 2~(a), 1.0~(b), 0.8~(c), and 0.5~$\mu$m (d). Scale bars are 2~$\mu$m. The lower right insets: Fourier spectra of the SEM images. The lower left insets: enlarged SEM images of typical nanoparticles from the arrays.}\label{2DSEM}
\end{figure*}

First, strongly focused fs-laser beam irradiates a thin Au film (Fig~\ref{scheme}a) at fluence slightly higher than the thin film ablation threshold, which increases in range of 15--70~mJ/cm$^{2}$ with growth of the film thickness. These values are much lower than the values for single-shot ablation thresholds of thin gold films of corresponding thicknesses~\cite{kuznetsov2009, MakarovOL2015}. Since the scanning conditions correspond to the number of pulses \emph{N}$\sim10^{4}$ and the time delay between the fs-laser pulses in train is 12.5~ns, the temperature in the vicinity of irradiated area is gradually accumulated pulse-by-pulse. Moreover, heating of free electrons in the irradiated area up to temperatures of 1~eV at intensities near~$1\times10^{12}$~W/cm$^{2}$~\cite{wang1994time} combined with the multiphoton photoexcitation makes it possible to launch considerable emission of electrons, pulling charged ions away from the surface~\cite{schmidt2000dynamics, ionin2014electron}. Therefore, multipulse gentle material removal takes place mainly in the directly irradiated region, while the surrounded area is heated via residual heat transfer. This mechanism is supported by our measurements (for details, see \textit{Methods} and Fig.S2 in \textit{Supplementary Materials}) of scale of laser energy deposition (0.64~$\mu$m), which is almost equal to the diameter of the focused laser beam (0.68~$\mu$m), being an advantage of the femtosecond laser pulse duration. Importantly, there is almost no ablation debris nearby each groove as compared to ns-laser ablation~\cite{kaganovskii2006periodic} or thermal regimes of fs-laser ablation~\cite{zhigilei2009ablation}, which are followed by random dissemination of nanoparticles in the vicinity of heated area. In our regime, rather clean and narrow (width of $\Delta\approx$400~nm$\leq$$\lambda$) grooves can be written directly on the surface (Fig.~\ref{scheme}, Fig.S3 in \textit{Supplementary Materials}). It is worth noting that femtosecond pulse duration minimizes the heat affected zone~\cite{bauerle2011laser}, resulting in a shallow groove depth up to 30$\%$ deeper than the film thickness at the highest fluences.

Second, a single element can be easily cut (Fig.~\ref{scheme}b) to a patch with the size larger than the uncertainty of the grooves edges ($\pm50$ nm, see Fig.~\ref{2DSEM}a). The cut patch is thermally isolated since the thermal conductivities of the silica substrate and air are two and four orders of magnitude smaller than gold, respectively. Therefore, the isolated patch can be easier heated up to the temperatures where the film undergoes dewetting process.~The dewetting temperature of the films could be much lower than the melting point of bulk gold 1337~K~\cite{bauerle2011laser}. For instance, a 10-nm Au film on fused silica is dewetted even at 430~K, while 60-nm Au film undergoes dewetting at temperatures lower than 870~K~\cite{gadkari2005comparison}. Moreover, for sub-picosecond laser pulses, when the pulse duration is shorter than the time of the pressure relaxation (picosecond scale for metals), the laser heating is followed by a high pressure waves excitation~\cite{zhigilei2000microscopic}, disturbing surrounding area, which might decrease the dewetting temperature~\cite{wensink2002dewetting}. Additionally, the smaller the patches (with smaller total volume), the higher the temperatures can be achieved at fixed fluence. As a result, the heated cut Au patch transforms into a nanoparticle of the same volume during the dewetting process (Fig.~\ref{scheme}c).

Although dewetting is a spontaneous process for homogeneous film~\cite{Bischof96, Trice07}, the reproducible formation of a certain number of nanoparticles has been demonstrated by heating of lithographically cut microscale patches of thin film~\cite{Lian06,Kim09APL,fowlkes2011self,wu2014directed,Bobod2014wafer}. Moreover, a single nanoparticle formation from each patch is possible, when patch width-to-height ratio ($\xi$=\emph{w/h}, where $w$ is the lateral size of the square patch and $h$ is the film thickness) is smaller than a certain value~\cite{Kim09APL}. In the case of Au film on SiO$_{\rm 2}$ substrate, this value must fulfill the condition $\xi$$<$40, to provide almost 100 percentage probability of single particle formation with a certain diameter~\cite{Kim09APL}. In our experiments we achieve the most controllable formation of nanoparticles at $\xi\approx$15--30.

This technique enables direct writing of 0D (a single nanoparticle, Fig.~\ref{scheme}d), 1D (a chain of nanoparticles, Fig.S3b in \textit{Supplementary Materials}) and 2D (an array of nanoparticles, Fig.S3c in \textit{Supplementary Materials}) nanoparticle-based structures. The 2D structures are produced by laser beam scanning of the film surface in two orthogonal directions, allowing extremely high production rate $\sim$~1~mm$^{2}$/min at available speed of laser scanning $\sim$~0.3~m/s. All types of the structures have been successfully fabricated on 20 and 30~nm Au films at large-scale (Fig.S4 in \textit{Supplementary Materials}). Beside the film thickness, another parameter that affects the diameter and quality of the nanoparticles is the width $w$ of the cut patch, which is determined by scan period (\textit{P}). Moreover, varying both scan period and film thickness, one can control type of 2D nanoparticles pattern from completely ordered to multiscale or disordered state.

\begin{figure}[t!]
\begin{center}
\includegraphics[width=0.45\textwidth]{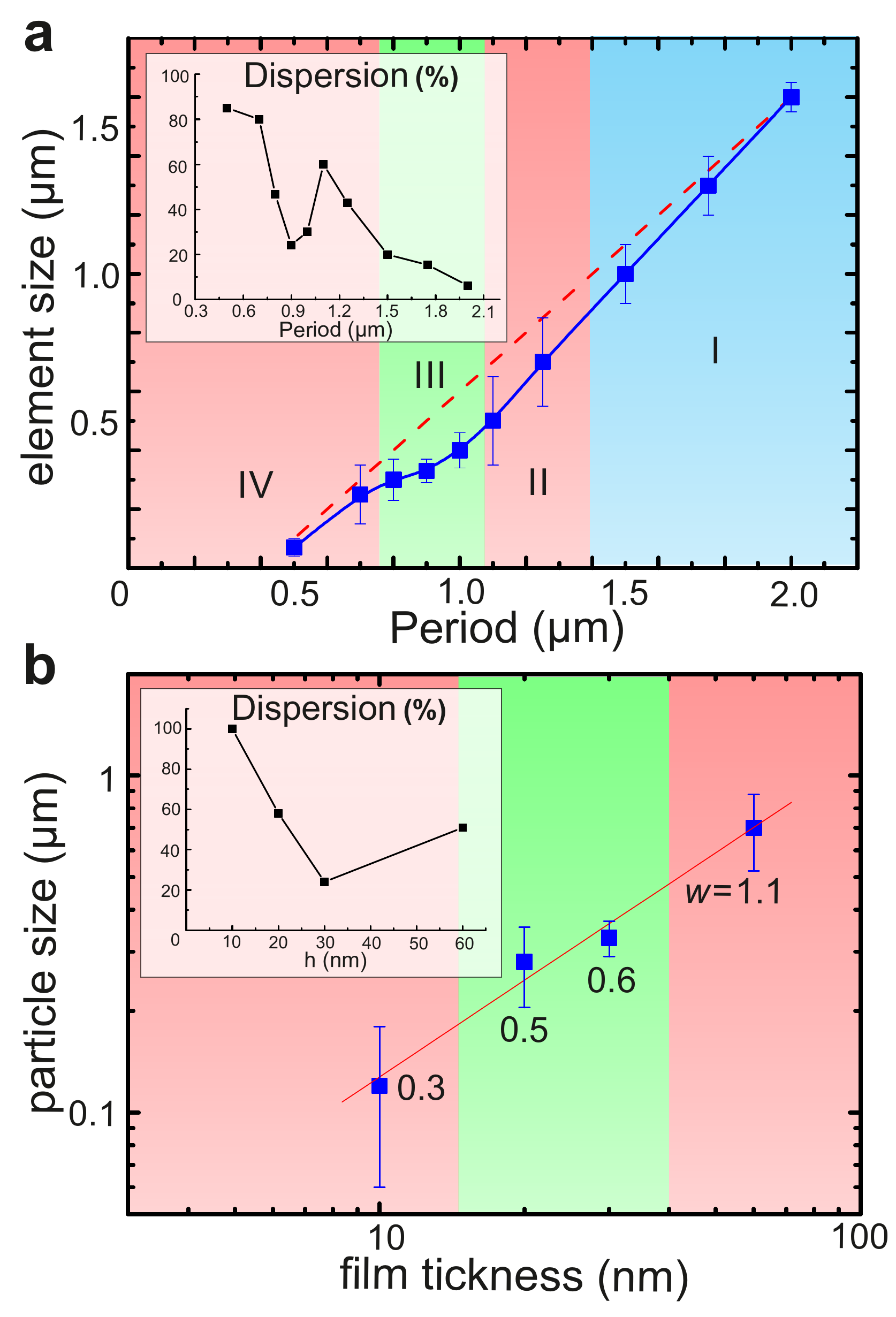}
\end{center}
\caption{(a) Measured dependence of a mean size of laser cut element on 30-nm Au film (blue squares) and the expected sizes calculated as the difference of the periods from groove width (red broken line). The inset represents dispersion of the size distribution as a function of the period of the crossed fs-laser scan. (b) Measured dependence of mean Au particle size on the film thickness at a given $w$. The inset represents dispersion of size distributions as a function of the film thickness. Red, green, and blue zones denote low, high, and moderate periodicity of the nanoparticles arrays, respectively.}\label{Dep}
\end{figure}

In Fig.~\ref{2DSEM}, SEM images of the fabricated 2D structures from a 30-nm Au film are shown as a sequence of reduced period between $90^{\circ}$-crossed laser scans at fluence of 40~mJ/cm$^{2}$. The laser fluence is chosen to provide as narrow and reproducible as possible grooves (Fig.~\ref{2DSEM}a). Evidently, there are four main regimes: (I) cutting of a square micro-patch (Fig.~\ref{2DSEM}a); (II) unstable transition from a square micro-patch to a single particle; (III) controllable cutting of a single spherical nanoparticles (Fig.~\ref{2DSEM}b); (IV) formation of more than one nanoparticles from the cut patch (Fig.~\ref{2DSEM}c,d).~Corresponding regimes are indicated in Fig.~\ref{Dep}a, where size-dispersion of cut patches and nanoparticles is a non-monotonic function of the scans period. In particular, for $h$=30~nm the size-dispersion demonstrates local minimum at scan the period of 0.9--1 $\mu$m, i.e.~in the most stable and reproducible regime of ordered nanoparticles formation for the 30-nm Au film.

For film thickness of $h=30$~nm, we observe reproducible formation of multiscale nanoparticles agglomerates (so-called "nanogalaxies"~\cite{gopinath2009plasmonic}) nearby each ordered nanoparticle when periods of scan are smaller than 0.9~$\mu$m (Fig.~\ref{2DSEM}c). Similar behaviour has been observed when some parts of molten film are not absorbed by larger particles during dewetting process, and interpreted in terms of multimodal Rayleigh-Plateau instability~\cite{fowlkes2014hierarchical}. Indeed, the agglomerates are formed preferably from the side of last cutting of each patch. Further reduction of scan period leads to fragmentation of large nanoparticles to smaller ones, yielding completely disordered pattern ("random" regime in Fig.~\ref{2DSEM}d).

The experimental dependence of the nanoparticles diameters on the film thickness exhibits almost linear (slope is 0.9) behavior over the whole range (Fig.~\ref{Dep}b), being a general trend for thin film dewetting~\cite{thompson2012solid}. As it was mentioned before, we observe a transformation of each patch into a single nanoparticle, which is due to the too small sizes of the patches in comparison with typical wavelength of Rayleigh-type instability~\cite{Kim09APL}. Since the nanoparticles are embedded into silica surface (Fig.~\ref{Opt1}a), the dewetting process occurs at temperatures close to fused silica melting point 1873~K~\cite{bauerle2011laser}, exceeding the melting temperature of gold 1337~K~\cite{bauerle2011laser}. If we assume that dewetting of gold film occurs in liquid state, we can consider the value of wavelength of the dominating capillary wave on homogeneous film, which is estimated as~\cite{wyart1990drying}: $\lambda_{\rm m}=2\pi\sqrt{2/3} h^{2}/\alpha$, with the parameter $\alpha$ being characteristic of the material combination substrate/film ($\alpha\approx$4.3~nm for gold on fused silica~\cite{Bischof96}). Substituting our experimental parameters, we obtain the following estimations: $\lambda_{\rm m}$($h$=10~nm)$\approx$0.12~$\mu$m, $\lambda_{\rm m}$($h$=20~nm)$\approx$0.48~$\mu$m, $\lambda_{\rm m}$($h$=30~nm)$\approx$1.1~$\mu$m, and $\lambda_{\rm m}$($h$=60~nm)$\approx$4.3~$\mu$m. Comparison of these values with corresponding values of $w$ plotted in Fig.~\ref{Dep}b shows that the optimal conditions for single nanoparticle formation can not be fulfilled only for $h$=10~nm, which provides $\lambda_{\rm m}<w$ and is comparable with the uncertainties ($\pm50$~nm) of grooves edges. Similar to the scan period (the patch width) dependencies, there are also some ranges of thicknesses where the nanoparticles are the most monodisperse (Fig~\ref{2DSEM}b). Non-monotonic behaviour of the size dispersion dependence on scan period and film thickness can be attributed to two main reasons: (i) the conditions $w/\lambda_{\rm m}$$>$1 and $\xi$$<$40 are not fulfilled for very small thicknesses and (ii) the dewetting temperature is much higher for thicker films.

Therefore, too thin films can be applied only for random pattern formation at relatively low temperatures, being a good regime for the substrates with some restrictions on high temperatures. Too thick films or large patches can be hardly heated homogeneously during laser cutting, and do not allow achieving controllable dewetting.

In the next section, we will show possible applications of controllable laser-induced dewetting of Au films into three main functional types of pattern: ``ordered", ``agglomerates", and ``random".

\section{Applications}

Gold nanoparticles of different sizes are well known to provide resonant light absorption/scattering and local field enhancement in the optical range. In order to show applicability of the novel method, we examine 2D nanoparticles arrays fabricated at different regimes on two types of substrates for resonant absorbtion/scattering, sensing, and photovoltaics.

\subsection{Resonant absorbtion and scattering}

\begin{figure}
\begin{center}
\includegraphics[width=0.4\textwidth]{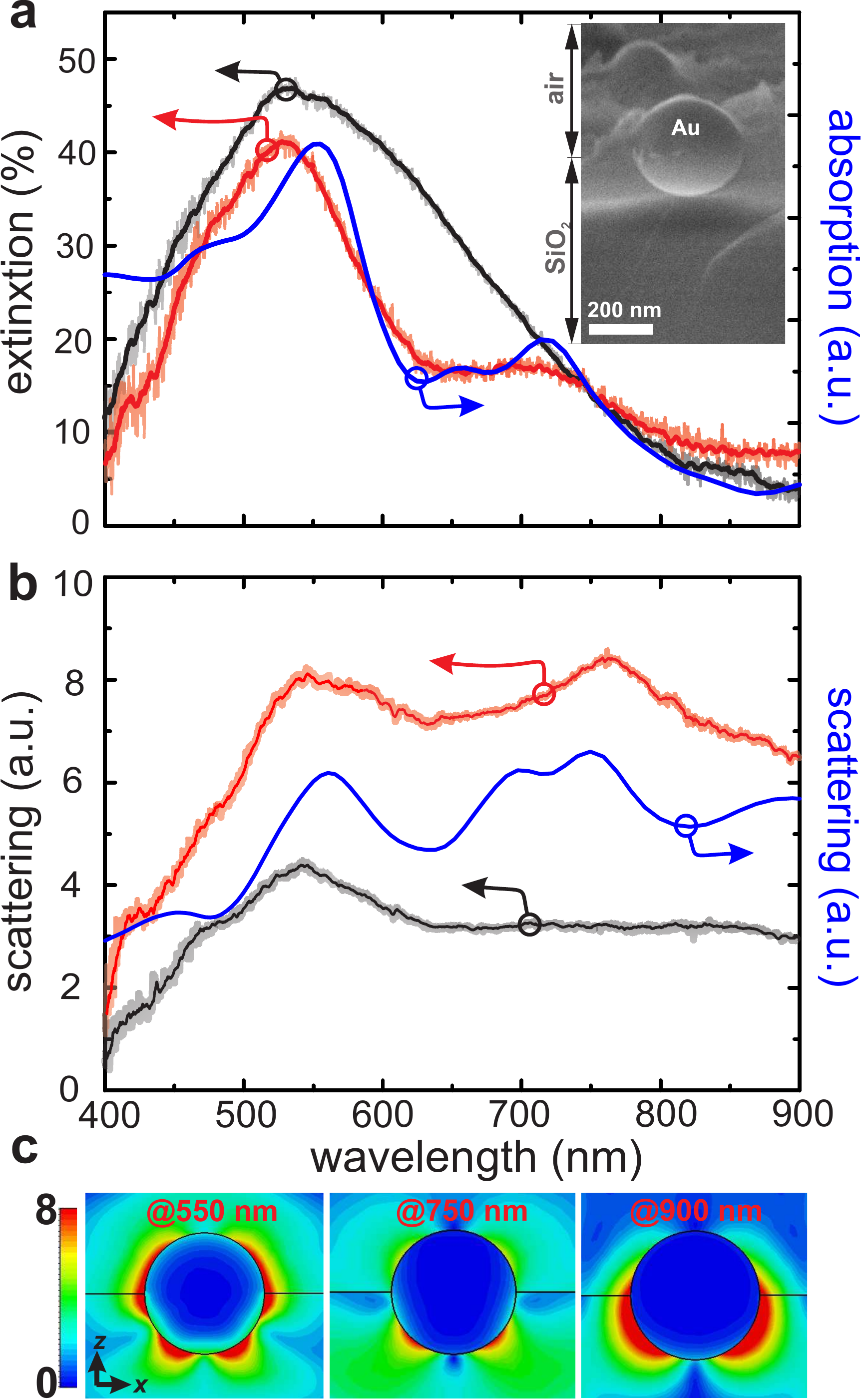}
\end{center}
\caption{Measured extinction (a) and scattering (b) spectra of ordered in 1$\times$1 $\mu$m scanning regime (red curve, left axis) and disordered in 0.7$\times$0.7~$\mu$m scanning regime (black curve, left axis) gold nanoparticles. Numerically calculated absorbtion (a) and scattering (b) cross-sections (blue curves, right axis) of a gold nanoparticles array (\textit{D}=330~nm and \textit{P}=1~$\mu$m). Inset in (a) shows cross-sectional SEM image of a single nanoparticle. (c) Numerically calculated distributions of the total normalized electric field intensity $|E|^{2}/|E_{\rm inc}|^{2}$ around the spherical gold particle, with \textit{D}=330~nm, irradiated by a plane light wave propagating in the negative z direction and linearly polarized along the $x$ axis.}\label{Opt1}
\end{figure}

The ordered 2D array of similar resonant metal nanoparticles is a key element for plasmonic metasurfaces and filters~\cite{maier2007plasmonics,grigorenko08extremely,boltasseva2008plasmonic,meta2014capasso,alu2014giant,meta2015kivshar}. In turn, multipolar (higher-order) plasmon resonances in relatively large nanoparticles have typically higher quality factors than the dipolar plasmon resonance and more flexible scattering diagram. The fabricated 330-nm particles in the ''monodisperse'' regime (III) have almost spherical shape and are partly embedded into the substrate (Fig.~\ref{Opt1}a), owing to dewetting on the molten substrate. The measurements of extinction (Fig.~\ref{Opt1}a) and scattering (Fig.~\ref{Opt1}b) spectra from ordered 330-nm nanoparticles array revealed three resonances at $\lambda$~$\approx$~550~nm, $\lambda$~$\approx$~750~nm and $\lambda$~$\approx$~900~nm (for the measurements details see \textit{Methods}). These resonances are in good qualitative agreement with our numerical simulation (Fig.~\ref{Opt1}a,b) for an array of gold 330-nm spheres embedded on half in fused silica infinite substrate (for calculation details see \textit{Methods}). According to our simulations of the near field distributions, these maxima correspond to octupole (550~nm), quadrupole (750~nm) and dipole (900~nm) plasmon resonances (Fig.~\ref{Opt1}c) in qualitative agreement with previous studies~\cite{evlyukhin2012optical}. Slight splitting of the resonances is attributed to the effect of substrate, inducing "image" of the multipolar plasmons under the nanoparticle~\cite{knight2009substrates}. It is worth noting that in comparison with ordered monodisperse nanoparticles fabricated under optimal conditions (\textit{h}=30~nm, \textit{P}=1~$\mu$m), the disordered nanoparticles (\textit{h}=30~nm, \textit{P}=0.7~$\mu$m, Fig.~\ref{2DSEM}c) with wide size distribution (from sub-100~nm up to 0.5~$\mu$m) exhibit single and broad extinction resonance. Indeed, sub-100~nm nanoparticles agglomerates surround each hundreds-nm particle (Fig.~\ref{Opt1}a), providing additional strong dipole resonance contribution into scattering/absorbtion spectra in the range of $\lambda$=500--600~nm.The random nanoparticles after the dewetting of the 10-nm Au film exhibit similar broad spectra.

\subsection{Sensing}

Developing of low-cost methods for fabrication of substrates for sensing applications is one of the most actively studying field of plasmonics~\cite{Halas2007nano, anker2008biosensing}, where cheap and environment-friendly laser technologies seem to be very promising~\cite{vorobyev2013direct}. Especially, it is desirable to achieve large-scale fabrication of plasmonic nanoresonators exhibiting extremely strong incident light localization and enhancement in the visible range.

\begin{figure}
\begin{center}
\includegraphics[width=0.5\textwidth]{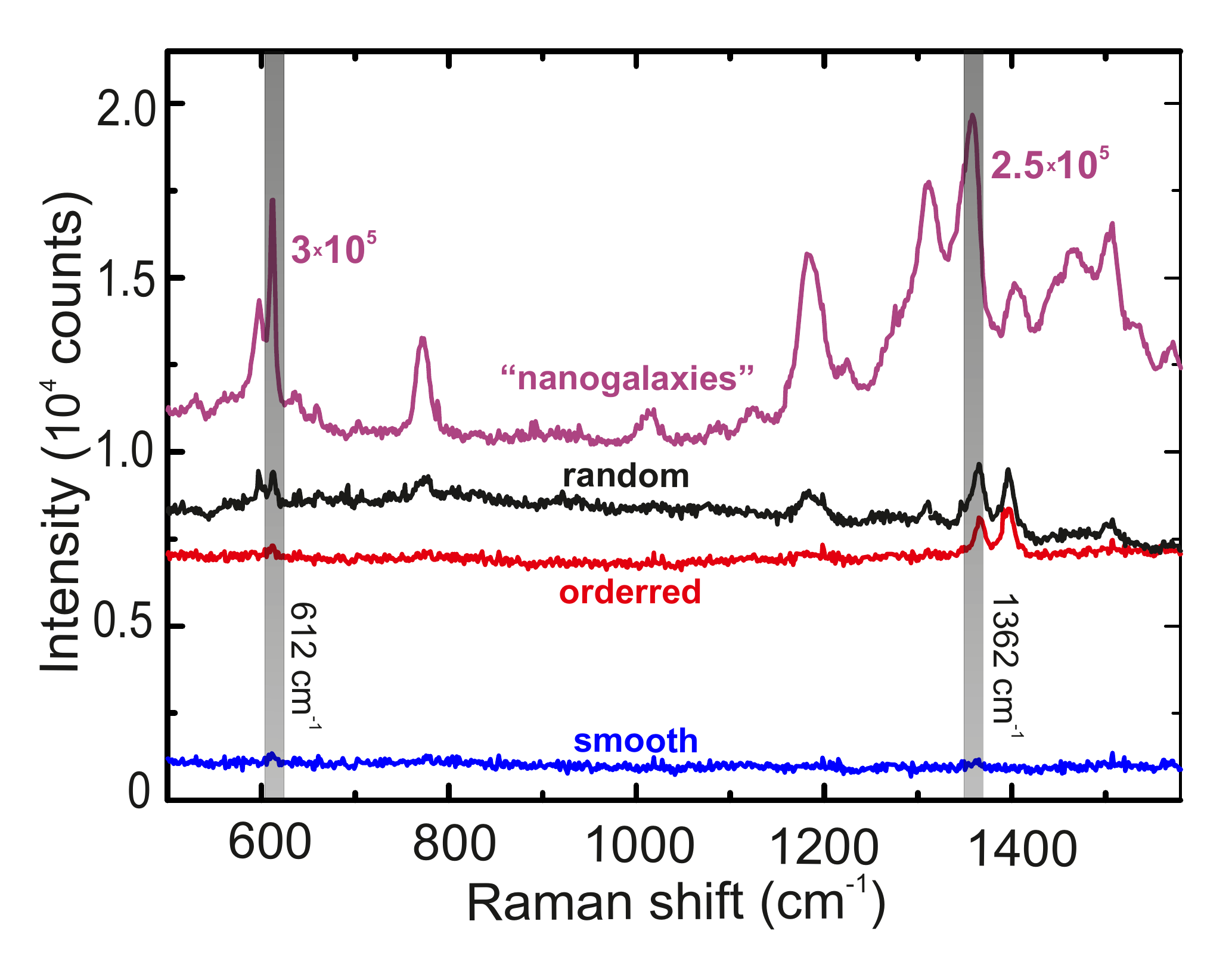}
\end{center}
\caption{(a) Raman scattering spectra from monolayer of R6G deposited on smooth (blue curve) and patterned 30-nm Au film at the fluence of 40~mJ/cm$^{\rm 2}$ and different scan periods: 1~$\mu$m (red curve), 0.8~$\mu$m (magenta curve), and 0.5~$\mu$m (black curve).}\label{Opt2}
\end{figure}

In Fig.~\ref{2DSEM}c one can see that direct laser writing provides formation of not only individual ordered nanoparticles, but also multiscale aperiodic nanostructures (``agglomerates"), which are more suitable for sensing applications~\cite{gopinath2009plasmonic}. In order to show their applicability, we measured Raman scattering signal from the deposited molecules of rhodamine-6G (R6G) on the nanostructured substrates (for the measurements details, see \textit{Methods}). The typical Raman scattering spectra from a single layer of R6G on different substrates with fs-laser fabricated Au nanoparticles are shown in Fig.~\ref{Opt2}. The systematic study of Surface-Enhanced Raman Scattering (SERS) signal on a scan periods reveals its strong dependence on the type of nanostructures. The relatively low average enhancement factor (EF) $EF<10^{5}$ corresponds to both micro-patches and ordered nanoparticles fabricated at cross-scan periods \emph{P}~$>$~0.7~$\mu$m, being approximately one order of magnitude higher than SERS signal from R6G on a smooth Au film (Fig.S5 in \textit{Supplementary Materials}). Nevertheless, these structures exhibit up to 5-fold higher $EF$ than that for a smooth Au 30-nm film with roughness amplitude less than 1 nm, which provides the so-called chemical mechanism of SERS and weak electromagnetic contribution on the roughness~\cite{moskovits1985surface}. The surface of the structures is not planar, and the main contribution to Raman signal gives R6G molecules from ordered Au nanoparticles with surface area of four times smaller than the area of smooth substrate surface within laser spot. Taking into account this fact, the SERS signal for the ordered array of the nanoparticles is about one order of magnitude larger in comparison with the signal on the film. Our numerical modeling of the electric field distribution nearby a single 330-nm Au nanoparticle embedded in fused silica shows, that the highest $EF$ value is lower than 100 in the hot spots, providing 3-fold electric field enhancement ($|E_{\rm loc}|/|E_{\rm inc}|$).

\begin{figure*}
\begin{center}
\includegraphics[width=0.95\textwidth]{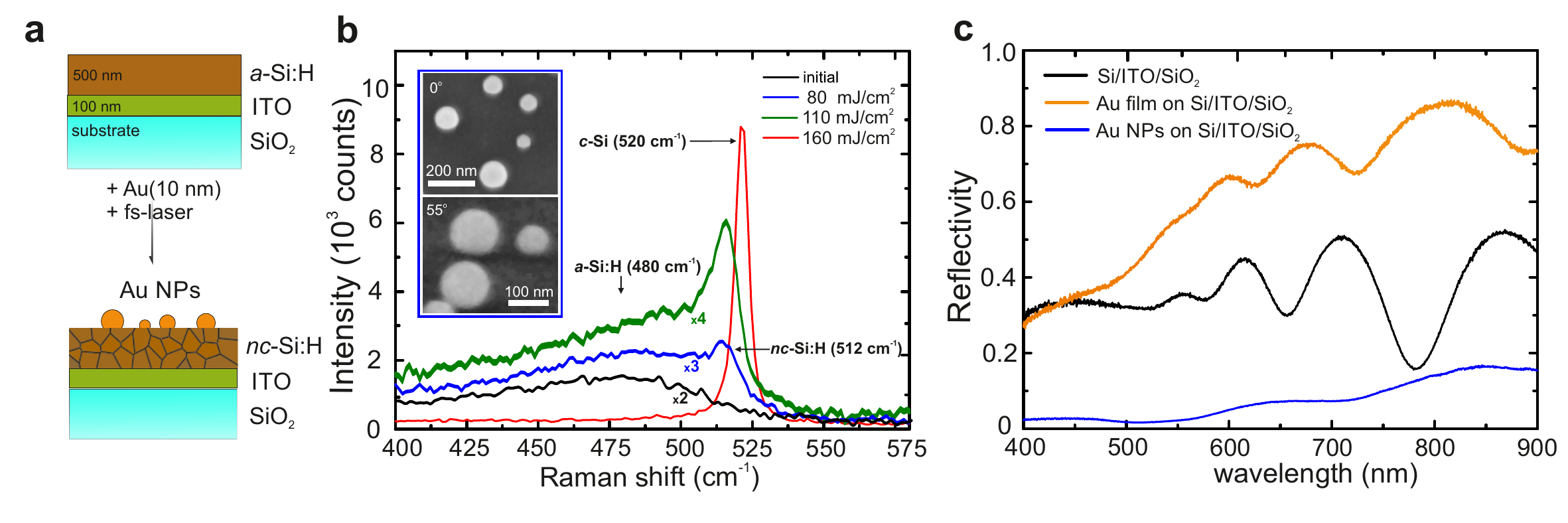}
\end{center}
\caption{(a) Schematic scketch of the process of solar element improving by fs-laser-induced dewetting of the deposited Au film.(b) Raman spectra from initial Si(500~nm)/ITO(100~nm)/SiO$_{2}$(substrate) multilayer structure (black curve) and from Au(10~nm)/Si/ITO/SiO$_{2}$ structure irradiated by 80~mJ/cm$^{2}$ (blue curve), 110~mJ/cm$^{2}$ (green curve), and 160~mJ/cm$^{2}$ (red curve). Inset: SEM images (top and 55$^{\circ}$ views) of Au nanoparticles fabricated at the fluence of 80~mJ/cm$^{2}$. (c) Measured at normal incidence reflectance from the Si/ITO/SiO$_{2}$ multilayer structure (black curve), Au/Si/ITO/SiO$_{2}$ (yellow curve) multilayer structure and laser fabricated Au nanoparticles on the Si/ITO/SiO$_{2}$ multilayer structure (blue curve). }\label{Trap}
\end{figure*}

The strongest average enhancement factor $EF > 10^{\rm 5}$ is observed on the structured 30-nm film at the scan periods of 0.7~$\mu$m and 0.8~$\mu$m, where extremely hot spots are formed in the gaps within agglomerates of sub-100~nm nanoparticles. The SEM images of the agglomerates reveal only $\sim$10$^{\rm 2}$ pairs of the sub-100-nm nanoparticles separated on a few nanometers nearby each 330-nm Au nanoparticle (Fig.~\ref{2DSEM}c). Therefore, one can assume that only $\sim$10$^{\rm 3}$~--~10$^{\rm 4}$ molecules from these gaps contribute to the average $EF$~$>$~10$^{\rm 5}$ over whole laser spot with $\approx$~1.1~$\times$~10$^{6}$ molecules, resulting in local $EF$ in the gaps up to $\sim$10$^{\rm 9}$--10$^{\rm 10}$. Indeed, our numerical calculations of two touching gold 20-nm nanoparticles provide electric-field maximum enhancement at $\lambda$=633~nm in the range of $|E_{\rm loc}|/|E_{\rm inc}|$$\approx$10--20 in place of fitting of a 1.5-nm R6G molecule between the particles, resulting in electric-field contribution to SERS local enhancement factor of $(|E_{\rm loc}|/|E_{\rm inc}|)^{4}$$\sim$10$^{5}$. Since the estimated chemical contribution to SERS from measurements of Raman signal of R6G on thin Au 30-nm film lies in the range of (5~--~9)~$\times$~10$^{\rm 3}$, the total \textit{EF} in the gap can achieve the level of $\sim$10$^{\rm 9}$, which is comparable with previous reports~\cite{pendry1996SERS, Halas2007gapSERS, ye2012plasmonic}.

\subsection{Photovoltaics}

Our method can be applied not only to glass substrates, but also to semiconductor and multilayer ones, i.e. to typical materials in solar industry. Metallic nanoparticles can be used as subwavelength elements to couple and trap solar radiation into an absorbing semiconductor thin film via both scattering and near-field enhancement mechanisms, eliminating reflection losses~\cite{atwater2010plasmonics, cui2014plasmonic}. This concept is particularly useful in thin-film solar cells~\cite{pillai2007surface, tan2012plasmonic, simovski2013enhanced}. However, beside relatively high reflectivity, thin-film solar cells, based on amorphous semiconductors (one of the most popular is \textit{a}-Si:H), suffer from very small carrier diffusion lengths, low absorbtion in IR, and short operation time due to light-induced degradation. To avoid these disadvantages, more complicated methods for nanocrystalline thin layer deposition are used~\cite{butte2000structural}. Simple thermal Au dewetting and \textit{a}-Si:H annealing are not safe, because heating of whole solar cell at several hundreds degrees can initiate interlayer diffusion. Therefore, the cost-effective method, combining fabrication of antireflection coating and gentle nanocrystallisation, is required.

In our experiments, the initial substrate for Au nanoparticles represents a typical multilayer structure for thin-film solar cells~--~a 1~$\times$~1~cm$^{2}$ sample of Si(500 nm)/ITO(100 nm)/SiO$_{2}$(bulk), where the silicon layer was deposited by the plasma enhanced chemical vapour deposition (PCVD) of SiH$_{3}$ gas (Fig.~\ref{Trap}a). The PCVD method provides deposition of an amorphous hydrogenated silicon layer (\textit{a}-Si:H) with typical broad peak in Raman spectrum around 480~cm$^{-1}$ (Fig.~\ref{Trap}b). The multilayer structure reflects about 30$\pm$10~$\%$ in the visible range with existence of Fabry-Perot-like resonances (Fig.~\ref{Trap}c).

Since laser-induced dewetting of 10-nm thick Au film results in the broadest extinction spectrum (Fig.~\ref{Opt1}a), whereas its dewetting temperature and threshold fluence are lower compared to the thicker films, we use the relatively thin (10-nm) Au film. In order to find the regime of partial crystallisation of the initial \textit{a}-Si:H phase to $nc$-Si:H with Raman peak position at 512~cm$^{-1}$, fs-laser irradiation of the Au(10~nm)/Si(500~nm)/ITO(100~nm)/SiO$_{2}$(bulk) multilayer structure was carried out at different fluences and the scan speed of 1~mm/s. The most appropriate fluence regime corresponds to $\approx$~80~mJ/cm$^{2}$, where percentage of a $nc$-Si:H fraction ($X_{nc}$) in an amorphous matrix is about 20~$\%$.The value $X_{nc}$ is calculated form the following ratio: $X_{nc}~=~I_{nc}/(I_{nc}+0.8I_{a})$, where $I_{nc}$ and $I_{a}$ are the integrated intensities of Raman signal from $nc$-Si:H and \textit{a}-Si:H fractions, respectively. The ratio of the Raman diffusion cross-section for c-Si over that of a-Si:H is assumed to be 0.8~\cite{smit2003determining}. According to the Raman peak position at 512~cm$^{-1}$ and the spectral width of 20~cm$^{-1}$, a mean grain size is less than 5~nm~\cite{smit2003determining}. The crystalline fraction can be increased continuously up to 100~\% by increasing of incident fluence in the range of 80--160~mJ/cm$^{2}$ (Fig.~\ref{Trap}b) in agreement with previous reports~\cite{bauerle2011laser}. However, the exceeding of 80~mJ/cm$^{2}$ leads to deformation of the Si-layer due to accumulation of residual stresses after the crystallisation. Fs-laser ablation of the Au film with $h~>$~10~nm requires higher fluences due to higher thermal conductivity of \textit{a}-Si:H, which also causes surface deformation of the multilayer structure. In contrast, at the fulences lower than 80~mJ/cm$^{2}$ we observed neither deformations nor oxidation (Fig.~\ref{Trap}b), which usually occur after surface annealing by longer laser pulses~\cite{bauerle2011laser}.

The irradiation of the Au(10 nm)/Si(500 nm)/ITO(100 nm)/SiO$_{2}$(bulk) multilayer structure with optimized experimental conditions exhibits formation of random distributed Au nanoparticles with mean size of 110~nm with broad dispersion (Fig.~\ref{Trap}b), similarly to the dewetting on glass (Fig.~\ref{2DSEM}d). The nanoparticles surface density is about 20~$\mu m^{-2}$, which is comparable with the best samples with metal nanoparticles obtained by thermal dewetting, wet-chemistry and lithography~\cite{atwater2010plasmonics, cui2014plasmonic}. Such covering of the structure by the Au nanoparticles significantly reduces its reflectance down to sub-5~\% level in the most important spectral range of 400--600~nm with the lowest value of 1.5~\% in the vicinity of 520~nm (Fig.~\ref{Trap}c), being competitive with previous studies~\cite{atwater2010plasmonics, cui2014plasmonic}. The control measurements from the unmodified Au/Si/ITO/SiO$_{2}$ structure confirm that the antireflection is not related to simple presents of gold layer (Fig.~\ref{Trap}c), but caused by light scattering on the nanoparticles. As a result, we demonstrate for the first time, to the best of our knowledge, simultaneous formation of nanoparticles and gentle transformation of the amorphous substrate to nanocrystalline state for potential improving of plasmonic solar cells.

\section{Conclusion}

In conclusion, the developed novel single-step and lithography-free method of direct Au-nanoparticles fabrication opens new possibilities to create 0D, 1D, and 2D plasmonic nanostructures on large-scale. To the best of our knowledge, it is the first demonstration of simultaneous cutting and dewetting of thin metallic film that paves the way to extremely simplify the technology of monodisperse and ordered nanoparticles fabrication. Moreover, this method can be applied for dielectric, semiconductor and multilayer substrates, allowing to precisely control their microscopic properties. The systematic experimental study reveals several functional regimes of film patterning: ordered 2D monodisperse nanoparticles for resonant light absorption; "agglomerates" for sensing; and randomly distributed nanoparticles on semiconductor surface for antireflection. Additionally, the nanoparticles can be embedded on a half into the surface at higher laser fluences, providing optical near-field enhancement in the substrate material. We believe that the proposed method is a new stage in the rapidly growing field of ``large-scale nanophotonics``.

\section{Methods}
\textit{Fabrication.} A commercial femtosecond laser system (Femtosecond Oscillator TiF- 100F, Avesta Poject) was used, providing laser pulses at 780 nm central wavelength, with maximum pulse energy of 5 nJ, and pulse duration of 100 fs at the repetition rate of 80 MHz. Laser energy was varied and controlled by an acousto-optical modulator (R23080-3-LTD, Gooch and Housego) and a power meter (FielfMax II, Coherent), respectively, while the pulse duration was measured by an autocorrelator (Avesta Poject). Detailed experimental setup is presented in Supplementary Materials (Fig.S1).

Laser pulses were tightly focused by an oil immersion microscope objective (Carl Zeiss 100$\times$) with a numerical aperture (NA) of 1.4 (experimental setup is shown in Supplementary Information). According to the relation \emph{d}$\approx$1.22$\lambda$/NA, the estimated full-width at half-maximum diameter of the beam focal spot size is $d$=0.68~$\mu$m, which is close to measured value (0.64~$\mu$m) by standard method~\cite{liu1982simple} based on the dependence of laser-damaged area on incident laser energy (for details, see the Supplementary Information).

Laser beam was focused on supported Au films with thicknesses in the range of $h$=10--60 nm, thermally evaporated (Bock Edwards Auto 500) on the back side of 140-$\mu$m-thick silica glass without any additional adhesion-improving layers.~The samples were then placed on a three-dimensional air-bearing translating stage driven by brushless servomotors (ABL1000, Aerotech), allowing sample translation with various scan speeds up to 300~mm/s. In all experiments, film surface scanning by the laser beam was carried out with velocity of 5 mm/s, providing the number of laser pulses per each point of approximately $N\sim10^{4}$.

\textit{Characterizations}. Preliminary optical imaging of the structured films was provided immediately during the laser processing by integrated CCD camera, collecting transmitted through the film white light from a lamp (Fig.S1). The high-resolution morphology characterization was carried out by means of scanning electron microscope (SEM, Carl Zeiss, Neon 40) and atomic force microscope (AFM, SmartSPM AIST-NT). The Fourier spectra of SEM images were obtained in ImajeJ software.

Optical transmission (\textit{T}) and reflection (\textit{R}) broadband measurements were carried out at normal incidence of linearly polarized light from a halogen lamp (HL-2000-FHSA), using a commercial spectrometer (Horiba LabRam HR) with CCD camera (Andor DU 420A-OE 325). The excitation Olympus PlanFI (NA=0.95) and collection Mitutoyo M Plan APO NIR (NA=0.7) objectives were used for transmission measurements. Objective Mitutoyo M Plan APO NIR (NA=0.42) was employed for reflection measurements. Scattering (\textit{S}) spectra measurements were carried out in a dark-field scheme, where the arrays irradiation was performed at an angle of incidence of 70$^{\circ}$ with surface normal and scattered signal collection was performed by means of objective Mitutoyo M Plan APO NIR (NA=0.7). Confocal optical scheme was optimized for collection of all signals (\textit{T}, \textit{R} and \textit{S}).

\textit{Raman signal measurements and enhancement factor calculations.} The nanoparticles were functionalized with a selfassembled monolayer of R6G. The substrates were submerged in 0.5-mM solution made with water for 1 hour and then gently rinsed in neat water. Enhancement factor estimates were measured on a micro-Raman apparatus (Raman spectrometer HORIBA LabRam HR, AIST SmartSPM system). Using a 15-mW, 632.8-nm HeNe laser, spectra were recorded through the 50$\times$ microscope objective (NA=0.42) and projected onto a thermoelectrically cooled charge-coupled device (CCD, Andor DU 420A-OE 325) array using a 600-g/mm diffraction grating. Individual spectra were recorded from both single spots (1.8-$\mu$m diameter) on the substrates, and from a 1-cm thick cell of neat R6G for normalization. Acquisition times for measurements from R6G and solar cells were 1000~s and 30~s, respectively.

From these measurements, the averaged over the excitation beam size Raman scattering enhancement factor (\emph{EF}) for the substrate is estimated with $EF_{\rm exp} = (I_{\rm SERS}/I_{\rm norm}) (N_{\rm norm}/N_{\rm SERS})$, where $I_{\rm SERS}$ and $I_{\rm norm}$ are the intensities of a specific Raman band, respectively. Two Raman modes of R6G at 612 cm$^{-1}$ and 1360 cm$^{-1}$ were chosen for the \emph{EF} calculations. The number of probed molecules in the unenhanced sample of neat R6G, $N_{\rm norm}$, is approximated using the molecular weight and density of R6G and the effective interaction volume of the Gaussian laser beam with the neat R6G sample. To estimate the $N_{\rm norm}$, we calculated the effective excitation volume by using the equation: $V_{\rm ex} = \pi r^{2}H$, where \textit{r} is the radius of the beam size ($r$$\approx$0.9 $\mu$m) and \emph{H} is the effective depth of focus ($H$$\approx$9 $\mu$m). Thus, we estimated an effective excitation volume of 22.9 $\mu m^{2}$ for our Raman microscopy with 633 nm excitation using the objective. Then, $N_{\rm norm}$ was calculated by the expression: $N_{\rm norm} = C N_{\rm A}V_{\rm ex}\rho/M = 1.08\times10^{9}$ molecules, where $C$ is the concentration ($3 \% $), $\rho$ is the density of R6G (1.26 g/cm$^{\rm 3}$), \textit{M} is the molar mass of R6G (479 g/mol) and $N_{\rm A}$ is the Avogadro constant (6.02$\times$10$^{\rm 23}$ mol$^{-1}$). To determine $N_{\rm SERS}$, a self-assembled monolayer of R6G molecules (molecular footprint size of S$_{\rm R6G}$$\approx$2.2 nm$^{\rm 2}$~\cite{gupta2003high}) was assumed to be closely packed on the surface and the number of the molecules within focused beam can be estimated as \emph{N}=$\pi$$r$$^{2}$/S$_{\rm R6G}$$\approx$1.1$\times$10$^{6}$.

\textit{Numerical Simulations.} Numerical simulations are performed by using both frequency-domain and time-domain solvers in the commercial software CST Microwave Studio. Incidence of a plane wave at given wavelength on a gold~\cite{palik} sphere of a given diameter embedded on half in fused silica ($\varepsilon$=2.1) infinite substrate is considered.

\begin{acknowledgement}
This work was financially supported by Russian Science Foundation (Grant 15-19-00172).
The authors are thankful to Prof.~Y.~Kivshar, Dr.~S.~Kudryashov, Dr.~A.~Kuznetsov, and M.~Hasan for discussions.
\end{acknowledgement}

\bibliographystyle{nature}
\bibliography{AblationNew}

\end{document}